\documentclass{aa}
\usepackage{graphicx}
\usepackage{rotating}
\usepackage{placeins}
\begin{document}

\title{NGC\,3627: a galaxy-dwarf collision?$^*$}

\author{M. We\.zgowiec\inst{1}
\and M. Soida\inst{2}
\and D. J. Bomans\inst{1}}
\institute{Astronomisches Institut der Ruhr-Universit\"at Bochum, Universit\"atsstrasse 150, 44780 Bochum, Germany
\and Obserwatorium Astronomiczne Uniwersytetu Jagiello\'nskiego, ul. Orla 171, 30-244 Krak\'ow, Poland}
\offprints{M. We\.zgowiec}
\mail{mawez@astro.rub.de\\
$^{*}$Based on observations obtained with XMM-Newton, an ESA science mission with instruments and contributions
directly funded by ESA Member States and NASA}
\date{Received date/Accepted date}

\titlerunning{NGC 3627: a galaxy-dwarf collision?}
\authorrunning{M. We\.zgowiec et al.}

\abstract
{
Group galaxies very often show distinct signs of interaction with both companion galaxies and the intragroup medium. X-ray observations are particularly helpful because they provide information 
on the temperatures and the densities of the hot gas in galaxies and intergalactic space. This can put important constraints on the nature and timescales of these interactions.
}
{
We use the XMM-Newton X-ray observations of NGC\,3627 in the Leo Triplet galaxy group to explain peculiar features visible in the polarized radio maps.
}
{
We analyzed soft X-ray (0.2-1 keV) emission from NGC\,3627 to study the distribution of the hot gas and its temperature in different areas of the galaxy.
Any change throughout the disk can reflect distortions visible in the radio polarized emission. 
We also studied two bright point sources that are probably tightly linked to the evolution of the galaxy.
}
{
We find an increase in the temperature of the hot gas in the area of the polarized radio ridge in the western arm of the galaxy. 
In the eastern part of the disk we find two ultra-luminous X-ray sources. We note a large hot gas temperature difference (by a factor of 2) between the two bar ends.
} 
{
The polarized radio ridge in the western arm of NGC\,3627 is most likely formed by ram-pressure effects caused by the movement of the galaxy through the intragroup medium.
To explain the distortions visible in the eastern part of the disk in polarized radio maps, the asymmetry of the bar, and the distortion of the eastern arm, we 
propose a recent collision of NGC\,3627 with a dwarf companion galaxy. 
}
\maketitle

\section{Introduction}

NGC\,3627 (M66) is a barred galaxy of SAb type at a distance of 11\,Mpc (see Table~\ref{astrdat}) 
in the Leo Triplet galaxy group. The dynamical model of Rots (\cite{rots}) that assumes a relatively recent tidal 
interaction with another galaxy of the Leo Triplet, NGC\,3628, is widely accepted to describe interactions 
of the galaxies within this group of galaxies (e.g. Haynes et al. \cite{haynes79}). 
The proposed model of interaction of NGC\,3627 with NGC\,3628 agrees with the radio continuum emission
properties studied by Soida et al. (\cite{soida01}): they found characteristic signatures of an interaction, such as a steep gradient of the
total power emission on one (western) side of the galaxy, and a much flatter
gradient on the opposite (eastern) side. A very bright polarized radio emission ridge
detected on the western side resembles the compression of the interstellar medium (ISM) caused
by the interaction. In addition to these well-understood signs of interaction,
Soida et al. (\cite{soida01}) discovered the so far unique polarized 
intensity (magnetic) arm, which ignores the underlying optical spiral arm structure
and crosses it at a large angle. This unusual feature still remains
unexplained.

NGC\,3627 has been studied in a wide range of spectroscopic observations --
in \ion{H}{i} line emission (Zhang et al. \cite{zhang}), CO (e.g. Paladino et
al. \cite{paladino}; Dumke et al. \cite{dumke}), \ion{H}{$\alpha$}
(Chemin et al. \cite{chemin}). All authors reported a higher velocity dispersion
measured in the vicinity of the southern bar end with respect to the northern one.
High spectral resolution observations (in CO lines) showed a double line structure
close to the southern bar end, at the part of NGC\,3627 where the unusual
magnetic arm was discovered. The most recent CO line analysis (Watanabe at al.
\cite{watanabe}) reports no particular difference in the star formation rates
at the two bar ends of NGC\,3627.

To understand the differences in the ISM properties at the two bar
ends we performed an analysis of the public archive X-ray data from XMM-Newton X-ray satellite
observations. Those data have already been published, but the authors focused mainly on the AGN studies in 
galaxy samples (Galbiati et al.~\cite{galbiati05}; Corral et al.~\cite{corral11}; Brightman et al.~\cite{brightman11}) or 
the construction of the XMM-Newton Slew Survey catalog (Saxton et al.~\cite{saxton08}).
We present here a thorough spatial and spectral analysis of the extended emission in the most distinct features of the galactic 
disk. Additionally, we used pipeline-processed archive observations from the Chandra X-ray Observatory to obtain 
accurate positions for selected point sources.

\section{Observations and data reduction}
\label{obsred}

\begin{table}[ht]
        \caption{\label{astrdat}Basic astronomical properties of NGC\,3627}
\centering
                \begin{tabular}{cccccc}
\hline\hline
Morphological type & SABb	\\
Inclination	   & 57\degr	\\
Position angle	   & 173\degr	\\
R.A.$_{2000}$	   & 11$^{\rm h}$20$^{\rm m}$15$^{\rm s}$	\\
Dec$_{2000}$	   & +12\degr 59\arcmin 30\arcsec	\\
Distance\tablefootmark{a} & 11\,Mpc	\\
\hline
\end{tabular}
\tablefoot{
All data except the distance are taken from HYPERLEDA database -- http://leda.univ-lyon1.fr -- see Paturel et al.~(\cite{leda}).\\
\tablefoottext{a}{Taken from Kanbur et al.~\cite{kanbur03}.}
}
\end{table}

\begin{table}[ht]
        \caption{\label{xdat}The characteristics of the X-ray observations of NGC\,3627}
\centering
                \begin{tabular}{cccccc}
\hline\hline
Obs ID			  	  & 0093641101   \\
column density nH\tablefootmark{a} & 1.99   \\
MOS filter	          	  & medium   \\
MOS obs. mode		  	  & FF   \\
pn filter                 	  & medium   \\
pn obs. mode		  	  & EF   \\
Total/clean pn time [ks]  	  & 5.98/5.88   \\
\hline
\end{tabular}
\tablefoot{
\tablefoottext{a}{Column density in [10$^{20}$ cm$^{-2}$] weighted average value after LAB Survey of Galactic \ion{H}{i} (Kalberla et al.~\cite{lab}).}
}
\end{table}

We used archive observations of NGC\,3627 performed by the XMM-Newton telescope on 26 May 2001.
Although the observation is short, good cosmic weather allowed almost the entire data to be used for the analysis (see Table~\ref{xdat}).
The data were processed using the SAS 11.0 package (Gabriel et al.~\cite{sas})
with standard reduction procedures. Following the routine of tasks $epchain$ and $emchain$,
event lists for two EPIC-MOS cameras (Turner et al.~\cite{turner}) and the EPIC-pn camera (Str\"uder et al.~\cite{strueder})
were obtained. Next, the event lists were carefully filtered for bad CCD
pixels and periods of intense radiation of high-energy background. The resulting lists were checked for the residual exsistence of soft proton flare contamination, which 
could influence the faint extended emission. To do that, we used a script\footnote{http://xmm2.esac.esa.int/external/xmm\_sw\_cal/\\background/epic\_scripts.shtml\#flare}
that performs calculations developed by Molendi et al.~(\cite{spcheck}).
We found that the event lists are only slightly contaminated, which should have no significant influence on the spectral analysis. This is also because we used not too sensitive 
archive data and expect that any slight influence of such contamination would be still within the errors of the derived parameters.  
The filtered event lists were used to produce images, background images, exposure maps (with and
without vignetting correction), masked for an acceptable detector area using the images
script\footnote{http://xmm.esac.esa.int/external/xmm\_science/\\gallery/utils/images.shtml}.
All images and maps were produced (with exposure correction) in four bands of 0.2 - 1 keV, 1 - 2 keV, 2 - 4.5 keV, 4.5 - 12 keV. 
The images were then combined into final EPIC images and smoothed with a Gaussian filter to achieve a resolution of 15$\arcsec$ FWHM. 
To allow a better comparison with the radio maps and to increase the sensitivity for the low surface-brightness diffuse emission, 
we used another type of smoothing of the original count images. This time the images were smoothed adaptively with a maximum smoothing scale of 30$\arcsec$ FWHM.
The rms values were obtained by averaging the emission over a large source-free area in the final map.
Next, the spectral analysis was performed. 
For creating spectra only the event list from the EPIC-pn camera was used because it offers the highest sensitivity in the soft energy band. However, only the emission above 
0.3\,keV was analyzed because the internal noise of the pn camera is far below this limit\footnote{http://xmm.esac.esa.int/external/xmm\_user\_support/\\documentation/uhb}. Although not crucial 
when combined with MOS cameras to produce images, the exclusion of the softest emission below 0.3\,keV is important to obtain reliable good quality spectra.
Unsmoothed images for all bands were used to search for the background point sources using the standard SAS
{\it edetect\_chain} procedure. Regions found to include a possible point source were marked. The area was individually chosen
by eye for each source, to ensure exclusion of all pixels brighter than the surrounding background. Those areas were then used for construction of spectral regions, 
for which spectra were acquired. Similarly, the spectra of the background were obtained using blank sky event lists (see Carter \& Read~\cite{carter}). 
These blank sky event
lists were filtered using the same procedures as for the source event lists.
For each spectrum response matrices and effective area files were produced. For the latter, detector maps needed for extended emission analysis
were also created. The spectra were binned, which resulted in a better signal-to-noise ratio. To obtain a reasonable number of bins at the same time, we chose 
to have 25 total counts in one energy bin. Finally, the spectra were fitted using XSPEC~11 (Arnaud~\cite{xspec}).
We also used the XMM-Newton Optical Monitor data acquired during the same observations and produced an image in the UVW1 filter using the standard SAS {\it omchain} procedure.

\section{Results}
\label{results}

\subsection{Distribution of the X-ray emission}
\label{dist}

\begin{figure}[ht]
                        \resizebox{\hsize}{!}{\includegraphics{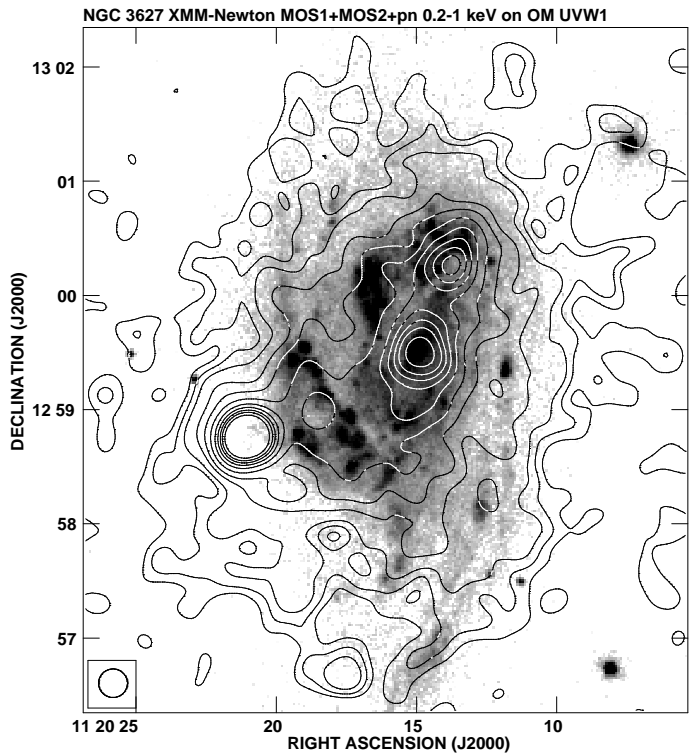}}
                \caption{Map of soft X-ray emission
                from NGC\,3627 in the 0.2 - 1 keV band overlaid onto the XMM-Newton Optical Monitor UVW1 filter image. The contours are 
		3, 5, 8, 16, 25, 40, 60, 80, 100, 120, 150 $\times$ rms. The map is convolved to the resolution of 15$\arcsec$. 
		The smoothing scale is shown in the bottom left corner of the figure.}
                \label{3627xfig}
        \end{figure}

The 15'' resolution image of NGC\,3627 (Fig.~\ref{3627xfig}) shows the hot gas emission following the optical disk of the galaxy. Howewer, in the lower part of the disk, an extension to the east is 
clearly visible. Within this extension, a bright point source can be seen, possibly an ultra-luminous X-ray source (ULX). 
The most interesting region, however, is the point-like source within the eastern arm, close to the southern end of the galactic bar. We present 
its spectral analysis and discuss its nature in Sect.~\ref{spectra}. We detected no X-ray emitting ridge in the western arm of the galaxy, where a polarized 
radio ridge, hence the more regular magnetic field, is visible.

\begin{figure}[ht]
                        \resizebox{\hsize}{!}{\includegraphics{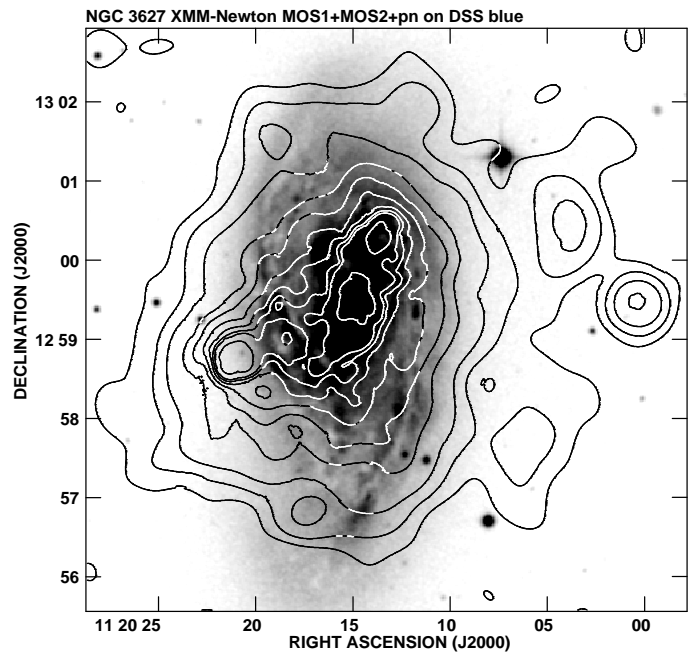}}
                \caption{Map of soft X-ray emission from NGC\,3627 in the 0.2 - 1 keV band overlaid onto the DSS blue image. 
		The contours are 3, 5, 8, 16, 25, 40, 60, 80, 100, 200 $\times$ rms. The map is adaptively smoothed with the largest scale of 30$\arcsec$.}
                \label{3627haloxfig}
        \end{figure}

\begin{figure}[ht]
                        \resizebox{\hsize}{!}{\includegraphics{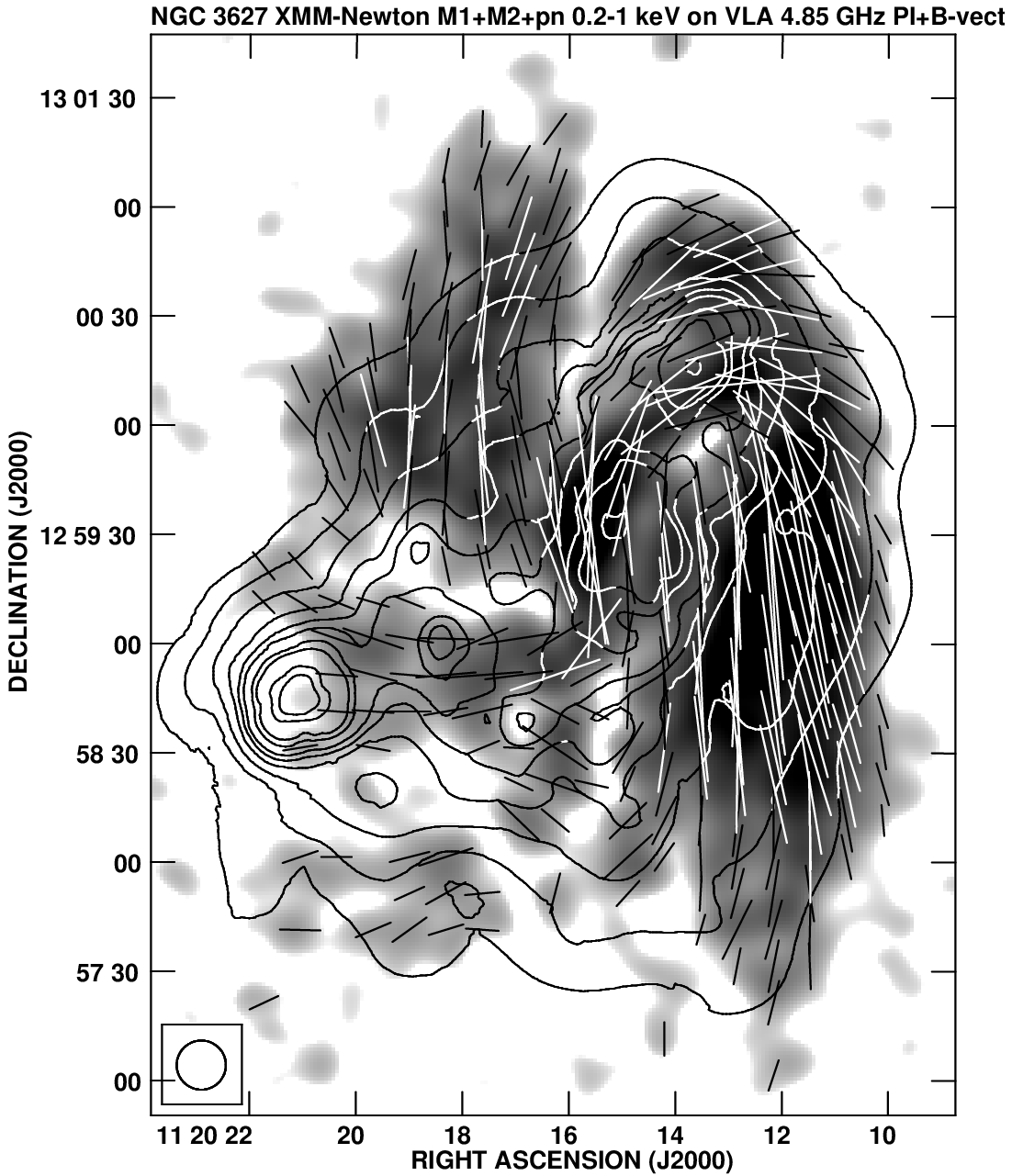}}
                \caption{Map of soft X-ray emission from NGC\,3627 in the 0.2 - 1 keV band overlaid onto the VLA radio polarization map at 4.85~GHz from Soida et al.~(\cite{soida01}).
                The contours are 16, 25, 40, 60, 80, 100, 150, 200, 500, 1000 $\times$ rms. The map is adaptively smoothed with the largest scale of 30$\arcsec$.
                The beam of the radio map is shown in the upper right corner of the figure. A magnetic field vector of a length of 1' corresponds to radio polarized intensity of 0.25~mJy/b.a..
                The gray scale radio polarized map is cut at the 3$\sigma$ level.}
                \label{3627pi6}
        \end{figure}

The adaptively smoothed image of NGC\,3627 (Fig.~\ref{3627haloxfig}) makes the faint emission clearly visible in the galactic outskirts. 
We see that a low surface-brightness envelope exists around the galaxy.
Unfortunately, the archive data are not sensitive enough to determine via spectral analysis
whether this emission comes from the hot galactic halo or the intergalactic medium (IGM). However, a slight asymmetry of the hot gas within the optical disk of the galaxy 
(Fig.~\ref{3627pi6}) may give only a hint that the latter case may be observed, because the asymmetry would then be caused by the ram-pressure interaction, which most likely 
is happening (see Sect.~\ref{diff}). We can trace the extension to the east along Dec$_{2000}=+12\degr 59\arcmin$, which corresponds 
well to the eastern magnetic arm of the galaxy.

\subsection{Spectral analysis of the X-ray emission}
\label{spectra}

\begin{figure*}[ht]
\begin{center}
\resizebox{7.6cm}{!}{\includegraphics{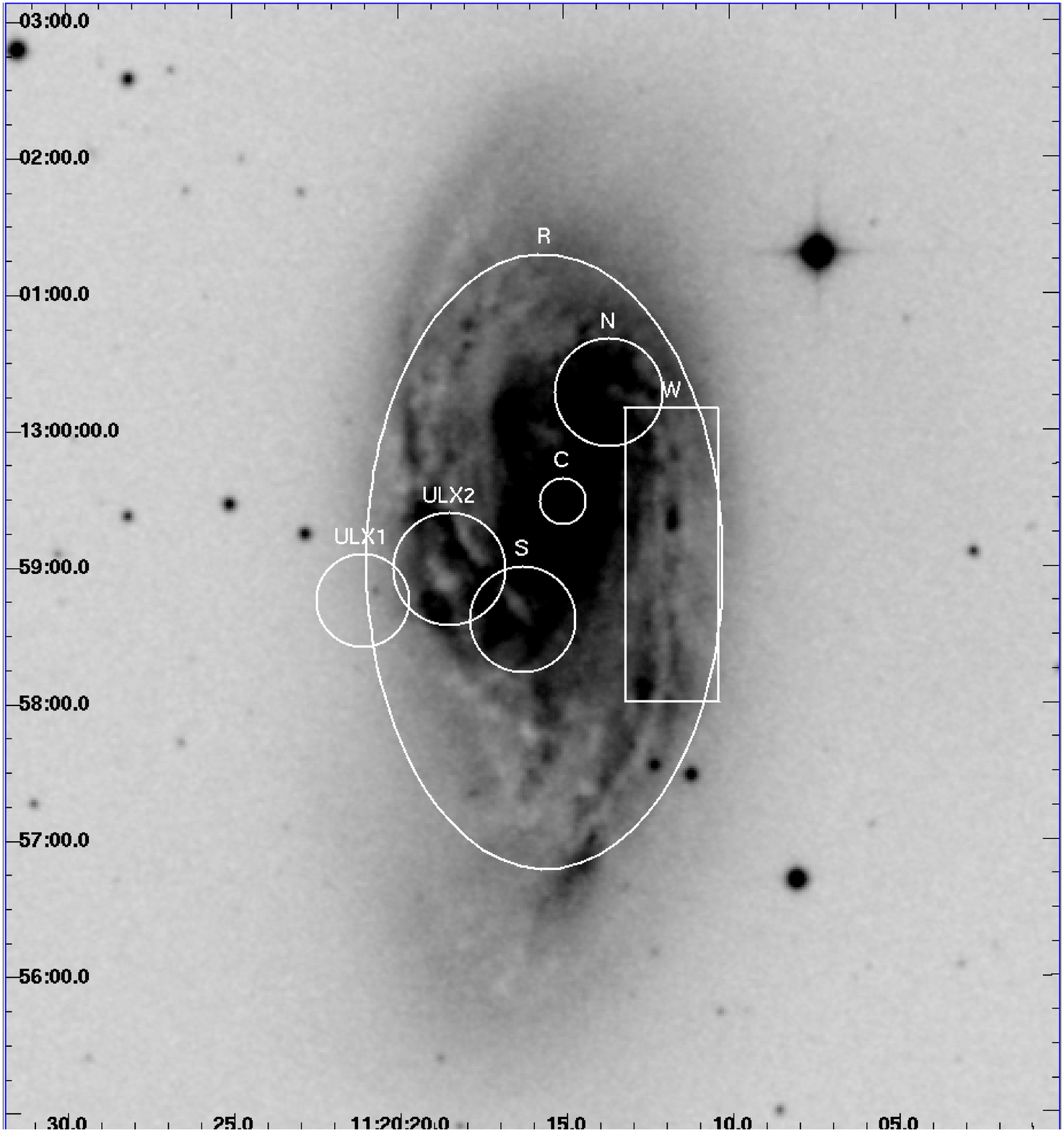}}
\resizebox{7.5cm}{!}{\includegraphics{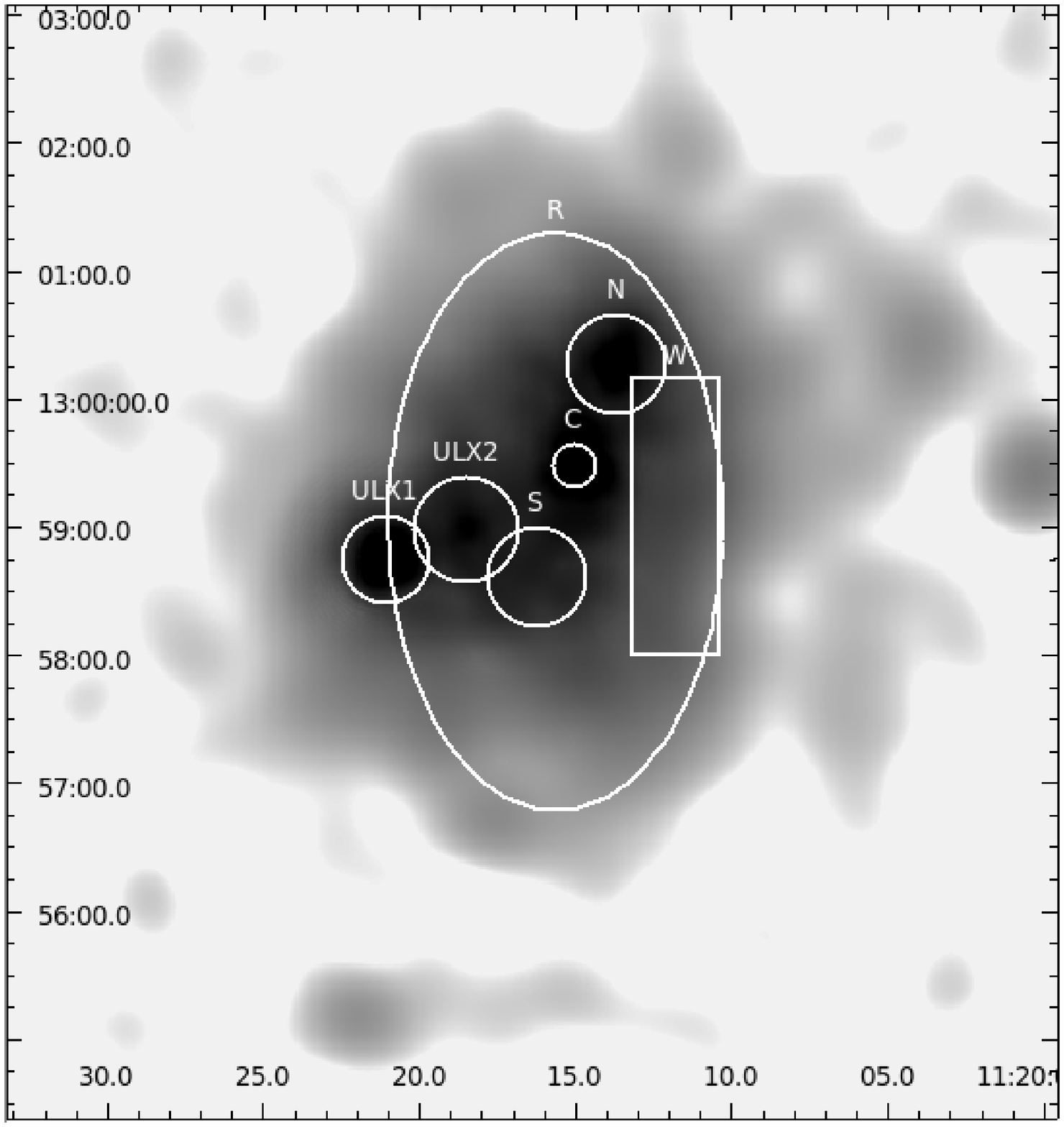}}
\end{center}
\caption{{\it Left}: Regions of NGC\,3627 for which the spectra were aquired (see text for a detailed description) overlaid onto the DSS blue image. 
	 {\it Right}: The same regions as in the left panel overlaid onto the map of soft X-ray emission in the 0.2 - 1 keV band as shown in Fig.~\ref{3627xfig}.} 
\label{3627xreg}
\end{figure*}

To examine the properties of hot gas and possibly the history of the interactions of the galaxy with the environment, we performed a spectral analysis of the selected features within 
the galaxy. All regions related to those features are presented in Fig.~\ref{3627xreg}. It includes the galactic nucleus (region C), the northern and southern end of the bar (regions N and S, resp.), 
a region corresponding to the radio polarized ridge in the western part of the galaxy (region W) and two point sources in the eastern part of the disk (regions ULX1 and ULX2).
To determine of the average temperature of the hot gas in the disk of NGC\,3627 we used the elliptical region (region R), which encompasses 
the entire X-ray emission from this galaxy, except for the regions mentioned above. 

For spectra of the hot gas investigated by us (all regions except ULX1 and ULX2) we used a model attributing to a thermal plasma and a contribution from still unresolved point-like sources.
Thermal plasma is represented in this work by a {\it mekal} model, which is a model of an emission spectrum from hot diffuse gas 
based on the model calculations of Mewe and Kaastra~(Mewe et al.~\cite{mewe}; Kaastra~\cite{kaastra}). In all models we fixed the metallicity to solar. 
A contribution from unresolved point-like sources is fitted with a simple power law.
For the regions of ULX1 and ULX2 a power law model was used. Since the region ULX2 still lies within a spiral arm of the galaxy, an additional component accounting for 
the internal absorbtion was used. For all models we also used a fixed foreground (galactic) absorption (see Table~\ref{xdat}). The errors provided for all model parameters (Table~\ref{3627xtab}) 
are always 1-$\sigma$ errors.

Both the central region and the southern end of the bar (regions C and S in Fig.~\ref{3627xreg}) show a similar temperature of the hot gas (0.26$^{+0.11}_{-0.06}$~keV and 
0.29$^{+0.09}_{-0.06}$~keV, resp.), which corresponds to the 
temperature of the hot gas throughout the whole disk of NGC\,3627 (0.29$\pm$0.04~keV, region R in Fig.~\ref{3627xreg}). 
A remarkable difference of a temperature of the hot gas can be seen in the northern end of the bar (region N). The gas is as hot as 0.71$^{+0.13}_{-0.12}$~keV in this region. This is almost twice as 
high as in other parts of the galaxy, except for the region W, where a radio polarized ridge is visible. The hot gas in this region has a temperature of 0.51$^{+0.12}_{-0.16}$~keV.

For the ULX1 and ULX2 sources we obtained photon indices of $\Gamma$ = 1.62$\pm$0.09 and $\Gamma$ = 2.01$^{+0.36}_{-0.27}$, respectively. The additional absorbtion in the model for the ULX2 source 
was fitted to be nH = 8.7$^{+6.9}_{-5.4}\times 10^{20}$\,cm$^{-2}$. 

\begin{figure*}[ht]
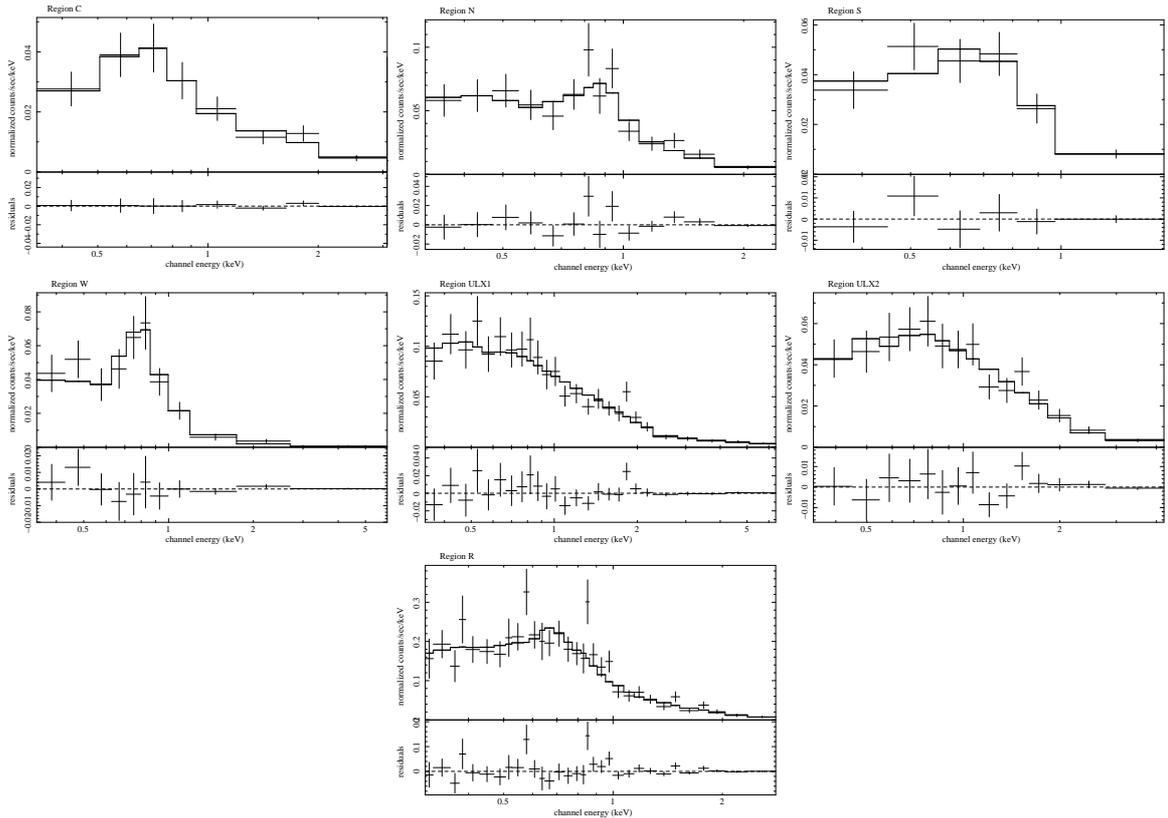

\begin{center}
\resizebox{5cm}{!}{\includegraphics[angle=-90]{3627_nucleus_new_normal.ps}}
\resizebox{5cm}{!}{\includegraphics[angle=-90]{3627_barN_normal.ps}}
\resizebox{5cm}{!}{\includegraphics[angle=-90]{3627_barS_normal.ps}}
\resizebox{5cm}{!}{\includegraphics[angle=-90]{3627_compression_normal.ps}}
\resizebox{5cm}{!}{\includegraphics[angle=-90]{3627_ulx_normal.ps}}
\resizebox{5cm}{!}{\includegraphics[angle=-90]{3627_clump_2wabs_normal.ps}}
\resizebox{5cm}{!}{\includegraphics[angle=-90]{3627_restemission_normal.ps}}
\end{center}
\caption{Model fits to the regions of NGC\,3627. See Tables~\ref{3627xtab} and
\ref{3627xf}.}
\label{3627mod}
\end{figure*}

All plots of the modeled spectra together with fitted models are presented in Fig.~\ref{3627mod}. 
The obtained parameters are presented in Table~\ref{3627xtab} and the derived X-ray fluxes are shown in Table~\ref{3627xf}.

\begin{table}[ht]
\caption{\label{3627xtab}Model-fit parameters of selected regions in NGC\,3627.}
\centering
\begin{tabular}{cccccc}
\hline\hline
Reg. 		  &	Internal	& kT 			   & Photon  			& Net	&	Red.     \\
\vspace{5pt} no.  &	nH\tablefootmark{a}	& [keV]  	   	   & Index         		& counts&	$\chi^2$ \\
\hline
\vspace{5pt}
C   		  &	--		& 0.26$^{+0.11}_{-0.06}$   & 1.09$^{+0.48}_{-0.28}$  	& 211	&	0.521    \\
\vspace{5pt}
N   		  &	--		& 0.71$^{+0.13}_{-0.12}$   & 2.18$^{+0.25}_{-0.26}$  	& 342	&	1.031    \\
\vspace{5pt}
S   		  &	--		& 0.29$^{+0.09}_{-0.06}$   & 2.37$^{+0.80}_{-0.98}$	& 181	&	1.033    \\
\vspace{5pt}
W   		  &	--		& 0.51$^{+0.12}_{-0.16}$   & 2.38$^{+0.60}_{-0.55}$	& 234	&	0.875    \\
\vspace{5pt}
R   		  &	--		& 0.29$\pm$0.04            & 2.09$^{+0.16}_{-0.19}$     & 1000	&	1.148    \\
\vspace{5pt}
ULX1              &                     & --                       & 1.62$\pm$0.09              & 725	&	0.741    \\
\vspace{5pt}
ULX2              &8.7$^{+6.9}_{-5.4}$  & --                       & 2.01$^{+0.36}_{-0.27}$     & 388	&	0.663    \\
\hline
\end{tabular}
\tablefoot{
\tablefoottext{a}{Column density in [10$^{20}$ cm$^{-2}$].}
}
\end{table}

\begin{table}[ht]
        \caption{\label{3627xf}Total (0.3 - 5 keV) and soft (0.3 - 1 keV) unabsorbed fluxes in 10$^{-14}$erg\,cm$^{-2}$s$^{-1}$ for modeled regions in NGC\,3627.}
\centering
\begin{tabular}{ccccc}
\vspace{5pt} Reg. no. & mekal & powerlaw  & total \\
\hline\hline
\vspace{5pt}
C (total)   	& 1.8$^{+1.3}_{-1.1}$  & 10$^{+9}_{-5}$ & 12$^{+5}_{-6}$ \\
C (soft)	& 1.5$^{+1}_{-0.8}$  & 1.7$^{+1.2}_{-0.7}$ & 3.2$^{+2.2}_{-1.6}$ \\
\hline
N (total)  	& 3$^{+1.6}_{-1.3}$  & 13$\pm$3 & 16$^{+5}_{-4}$  \\
N (soft)	& 2.3$^{+1.3}_{-1.2}$  & 6.1$^{+2.6}_{-1.9}$  & 8.4$^{+3.8}_{-3.1}$  \\	
\hline
S (total)  	& 2.3$^{+1.7}_{-1.2}$ & 3.7$^{+3.4}_{-1.3}$  & 6$^{+5}_{-2.5}$ \\
S (soft)   	& 2.1$^{+1.3}_{-1}$ & 2.1$^{+3.2}_{-1.5}$  & 4.2$^{+4.5}_{-2.5}$ \\
\hline
W (total)  	& 2.4$^{+1.1}_{-1}$ & 5.5$^{+1.6}_{-1.4}$  & 7.8$^{+2.7}_{-2.4}$ \\
W (soft)   	& 2.4$^{+0.3}_{-1.3}$ & 2.5$^{+3.4}_{-1.1}$  & 4.8$^{+3.8}_{-2.4}$ \\ 
\hline
R (total)  	& 8$^{+3.9}_{-2.7}$ & 31$^{+6}_{-4}$  & 39$^{+9}_{-7}$ \\
R (soft)	& 7.5$^{+3}_{-2.6}$ & 14$^{+4}_{-3}$  & 22$^{+7}_{-6}$ \\
\hline
ULX1 (total)	& -- 		     	 & 38$\pm$4 & 38$\pm$4 \\
ULX1 (soft)	& --			 & 12$\pm$1 & 12$\pm$1 \\
\hline
ULX2 (total)	& --  		     	 & 23$^{+10}_{-5}$ & 23$^{+10}_{-5}$  \\
ULX2 (soft)	& --			 & 10$^{+6}_{-3}$  & 10$^{+6}_{-3}$ \\ 
\hline
\end{tabular}
\end{table}

\section{Discussion}
\label{disc}

Because NGC\,3627 is believed to interact within the Leo Triplet galaxy group, we discuss results of the spectral analysis in terms of the influence of the group environment upon this galaxy. 
The hypothesis of a recent collision with a companion dwarf galaxy is also discussed. 

\subsection{Extended emission}
\label{diff}

The average temperature of the hot gas in the disk (region R) is 0.29$\pm$0.04~keV, which corresponds to a typical temperature for 
the cool phase of the hot ISM in a spiral galaxy (e.g. T\"ullmann et al.~\cite{tuellmann06}; Owen \& Warwick~\cite{owen09}). 
Similar temperatures can be found in the regions of the nucleus and the southern end of the galactic bar. 
However, a relatively low temperature for a bar end seems to be surprising. Usually, one expects a higher concentration of the \ion{H}{ii} regions 
and strong gas compressions due to the mass flow between the bar and the spiral arm. In this case, the temperature of the hot gas should also be higher, as we see 
in the northern end of the bar. To investigate this in more detail, we calculated gas densities in those regions using the parameters obtained from the spectral analysis.
To estimate the volumes of the regions, we assumed a disk thickness of 1\,kpc. This can still introduce uncertainties in the volumes by a factor of 2 and vary the values of derived 
gas densities by 30-50\%. For our derivation of densities we assumed fully ionized plasma (where n$_e\simeq nH$). 
For the fit to the whole galactic disk we obtained a density of 5.7$\times$10$^{-3}\times\eta^{-0.5}$\,cm$^{-3}$. 
The density of the hot gas for the northern and the southern end of the bar was found to be almost twice as high -- 11.1$\times$10$^{-3}\times\eta^{-0.5}$\,cm$^{-3}$ and 
10.9$\times$10$^{-3}\times\eta^{-0.5}$\,cm$^{-3}$, respectively. In the latter case, a temperature of 0.29$^{+0.09}_{-0.06}$~keV results in local pressure that is almost twice as high as 
in the galactic disk. For the northern part of the bar, however, a high temperature of the hot gas (0.71$^{+0.13}_{-0.12}$~keV) requires a significant rise (by a factor of $\simeq$5) 
of the pressure above the level observed for the whole galaxy. For both cases this seems to be justified by a significant (in the southern end of the bar) and a very high (in the northern end 
of the bar) local star formation activity, as seen in the UV and H$\alpha$ images (Figs.~\ref{3627xfig} and \ref{3627xfigmid}, respectively). However, UV and H$\alpha$ regions in the area of 
the southern end of the bar seem somewhat ''scattered'' along the beginning of the eastern spiral arm, 
which may be caused by tidal interaction. Such an encounter would also explain the visible N-S asymmetry of the bar, as well as a large 
difference of the temperatures of the hot gas between the two ends of the bar. In Sect.~\ref{dwarf} we present the hypothesis of a possible collision with a companion dwarf galaxy.
The nuclear region yields the highest
density of the hot gas (20.3$\times$10$^{-3}\times\eta^{-0.5}$\,cm$^{-3}$), which could explain its low temperature (0.26$^{+0.11}_{-0.06}$~keV) by a more effective radiative cooling.

For the western arm (region W), where a polarized radio ridge is visible, we calculated a local hot gas density of 6$\times$10$^{-3}\times\eta^{-0.5}$\,cm$^{-3}$, 
which is only slightly higher than the average value for the
galactic disk. Although the temperature derived from the model fit to this region is consistent within errors with the average temperature of the galactic disk,
this could be a hint of a local increase in the temperature of the hot gas, which might be attributed to the compression effects.
Assuming this is the case, one can expect a local pressure increase by a factor of $\simeq$1.8. This, 
together with a prominent dust lane well visible in optical images and lack of an enhanced star-forming activity in this region, suggests then the existence of local gas compresssions.

\subsection{Point-like sources}
\label{point}

\begin{figure}[ht]
                        \resizebox{\hsize}{!}{\includegraphics{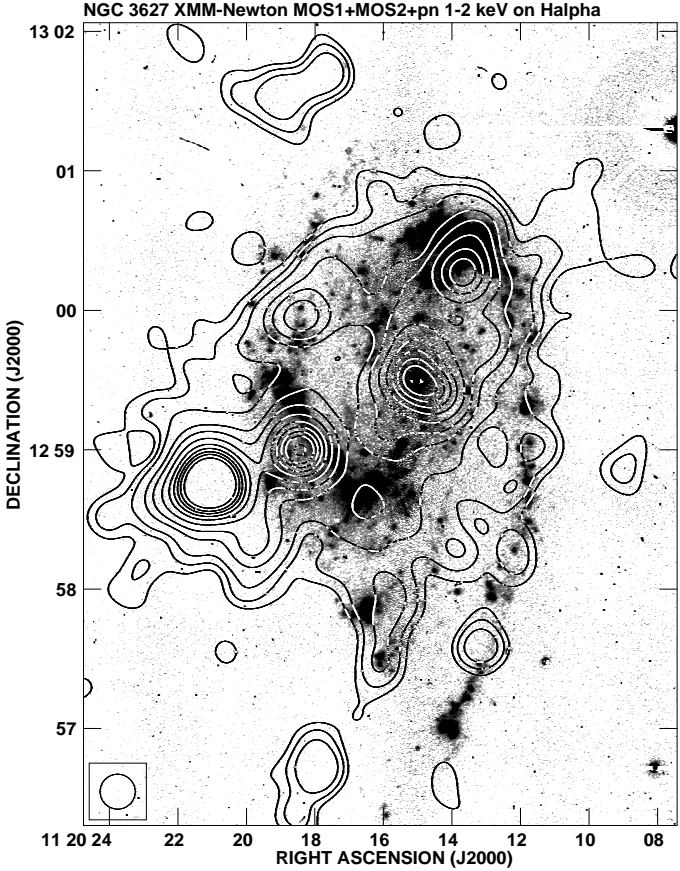}}
                \caption{Map of the X-ray emission
                from NGC\,3627 in the 1 - 2 keV band overlaid onto the
		H$\alpha$ image taken from the SINGS survey (Kennicutt et al.~\cite{sings}). The contours are
                3, 5, 8, 16, 25, 40, 60, 80, 100, 120, 150 $\times$ rms. The map is convolved to the resolution of 15$\arcsec$.
                The smoothing scale is shown in the bottom left corner of the figure.}
                \label{3627xfigmid}
        \end{figure}

In the vicinity of the eastern spiral arm of NGC\,3627, as mentioned in Sect.~\ref{dist}, two point-like sources are present. Those sources were also indentified by 
Swartz et al.~(\cite{swartz}) as ULXes. The spectral fits 
presented in Sect.~\ref{spectra} allowed estimates of the luminosities of those sources (see Table~\ref{ulxes}).
Comparing the fluxes (luminosities) of both sources we see that their values contradict their intensities, as observed in Fig.~\ref{3627xfig} - although source ULX2 is only $\simeq$1.6 times less 
luminous, it appears to be significantly fainter in the contour map of the soft (0.2-1~keV) X-ray emission. This clearly shows that the emission from this source is absorbed in the eastern spiral arm 
of NGC\,3627. Indeed, the fitted value of the internal absorption (8.7$^{+6.9}_{-5.4}\times 10^{20}$\,cm$^{-2}$) suggests that the softest X-ray emission is significantly diminished. 
A map in the higher energy band of 1-2~keV (Fig.~\ref{3627xfigmid}) confirms this by showing both sources to be of comparable intensity. We used archive Chandra observations
to verify the positions of both sources (see Table~\ref{ulxes}). For the ULX1 source we did not find any optical counterpart, which again confirms the nature of this source. 

For the ULX2 source, however, 
we may see a candidate for an optically emitting star. The probable source 
has a clear counterpart in the HST ACS images of NGC\,3627. After checking the HST astrometry against
the UCAC-3 catalog (Zacharias et al.~\cite{Zacharias2010}), the bright, isolated source coincides with the coordinates from the Chandra Source Catalog Release 1.1 (Evans et al.~\cite{evans10}, see 
Table~\ref{ulxes}). This release of the catalog already incorporated the correction discussed by Rots \& Budavari~(\cite{rotsbuda}). After correcting the HST
images to the coordinate system of the UCAC-3 catalog, the bright, extended source is located at R.A.$_{2000}$=11$^{\rm h}$20$^{\rm m}$18\fs33 Dec$_{2000}$=+12$\degr$59$\arcmin$00$\farcs$76. 
The Chandra position and the UCAC-3 reduced position on the HST ACS images agrees therefore within 0$\farcs$15. Since there is no other resolved source in a 1$\farcs$5 square box on the 
HST ASC images, we judge a chance superposition highly unlikely.

The ULX2 source on the ACS images is clearly extended with a FWHM of 3.7 pix compared to the FWHM of several similarly bright stars near the source, which
have a FWHM of 2.4 pix. The resulting PSF-corrected half-light radius of the
source is therefore 7\,pc, using a distance of 11\,Mpc to NGC\,3627. This size agrees perfectly with the typical size of young and old proto-globular
star clusters in, e.g., M\,33 and the Magellanic Clouds. Therefore, we can argue that the observed X-ray source may be a massive X-ray binary accreting at super-Eddington rate 
that resides within such a star cluster.
We measured the total magnitude of the cluster in the F435W, F555W, and
F814W and checked for excess emission in the F658W ACS frames. Following the recipes in the ACS section of the most recent HST data handbook and
correcting for galactic forground extincting based on Schlegel et al.~(\cite{schlegel98}), we derived total magnitudes for the cluster of B=23.299$\pm$0.008, V=22.785$\pm$0.008, 
I=22.009$\pm$0.008, resulting in B-V = 0.514 and V-I = 0.776 colors.  After subtracting the scaled F555W image no residual H$\alpha$ emission at the position of the cluster is visible on the F658N
image, hinting at the absence or small number of emission line stars in the cluster and no surrounding \ion{H}{ii} region. This may already hint at an age of the
cluster in excess of $1 \times 10^{7}$ years or that the cluster is massive enough that it has cleared the gas out of its environment. 
We can derive a more reliable age estimate, however, using the two measured colors of the cluster by comparing it to synthetic colors based on a STARBURST99 run using the
most recent code version (V\'azquez \& Leitherer~\cite{vazquez05}). We assumed the STARBURST99 standard parameters and solar metallicity. As a result, we estimate the
age of the cluster to be lower than $1.5\times10^{7}$ yr and the internal extinction on this line of sight in NGC\,3627 to be A$_V$ = 0.2 mag. 
Since the ULX2 source most likely resides within this cluster, we assume for this source the same age as for the cluster. 

Since both sources presented above are bright, they most likely still contribute to the spectral fit for the R region. This is because the sizes of their regions were chosen by eye (see 
Sect.\ref{obsred}) to include only the apparently significant emission. This in turn was performed to avoid excluding of too much data from the fit to the galactic disk region (R). Bright 
unresolved sources from regions C and N could also contribute to the region R. To estimate this contribution we used the SAS task {\em psfgen} to 
construct the PSF for this specific observation using the calibration data, as performed by Bogd\'an \& Gilfanov~(\cite{bogdan}). Next, using the actual sizes of our regions we 
calculated the energy fractions for all four sources. We found that the ULX1 source contributes with 7\% of its energy to the spectrum of the region R (we note that 
less than half of its region is included in region R), while for the ULX2 source it is 13\%. 
For the C and N regions the corresponding fractions are 40\% and 14\%, respectively. Using the fluxes for relevant regions (for regions N and R we used the fluxes of the power-law 
components) we calculated the total contribution from those sources to the spectrum of the region R to be 15.4$\times$10$^{-14}$erg\,cm$^{-2}$s$^{-1}$ and 4.6$\times$10$^{-14}$erg\,cm$^{-2}$s$^{-1}$ for 
the total and the soft flux respectively, which is 50\% and 33\% of the power-law component flux derived for the region R. Therefore, we conclude that a flux 
of 15.6$\times$10$^{-14}$erg\,cm$^{-2}$s$^{-1}$ still comes from unresolved point sources in the disk of NGC\,3627.

\begin{table*}[ht]
\caption{\label{ulxes}Positions and luminosities of the studied X-ray point sources in NGC\,3627.}
\centering
\begin{tabular}{cccc}
\hline\hline
Reg. & Chandra Source Name\tablefootmark{a} & Position\tablefootmark{a}          & L [erg\,s$^{-1}$]   \\
\hline
\vspace{5pt}
ULX1 & CXO J112020.9+125846 & R.A.$_{ICRS}$=11$^{\rm h}$20$^{\rm m}$20\fs90	 & 5.49$^{+0.62}_{-0.56}\times$10$^{39}$\\
     &  & Dec $_{ICRS}$=+12\degr 58\arcmin 46\farcs62 & \\
\vspace{5pt}
ULX2 & CXO J112018.3+255900 & R.A.$_{ICRS}$=11$^{\rm h}$20$^{\rm m}$18\fs32	&  3.35$^{+1.39}_{-0.78}\times$10$^{39}$\\
     &  	& Dec $_{ICRS}$=+12\degr 59\arcmin 00\farcs77 &	\\
\hline
\end{tabular}
\tablefoot{
\tablefoottext{a}{Taken from the Chandra Source Catalog Release 1.1 (Evans et al.~\cite{evans10}).}
}
\end{table*}  

\subsection{Is a recent collision with a dwarf galaxy probable?}
\label{dwarf}

NGC\,3627 shows clear marks of tidal and ram-pressure-induced distortions on
the western and eastern sides of its disk. The western part of the disk
is most likely ram-pressure-compressed, which suggests an extended polarized
radio ridge and a local increase of the temperature of the hot gas
(see Sect.~\ref{diff}). Furthermore, a slight asymmetry of the distribution of the hot gas, 
together with extensions of the hot gas to the east, along the magnetic arm (Fig.~\ref{3627pi6}),
can be a tracer of gas flows induced by the ram-pressure interaction. 
The asymmetry in the bar, together with the disturbance
in the eastern arm and the southern end of the bar speaks for the gravitational
origin of the disturbances of that area. The tidal interaction within the Leo Triplet
seems to be obvious, but according to the model by Rots (\cite{rots}),
the main encounter took place some 800\,Myr ago. The existence of 
massive binary systems (ULXes) may be caused by a local enhancement of the
star formation, triggered by a tidal distortion. However, recent interaction
with NGC\,3628 does not fit this scenario, because the age of the ULX2 source is of
the order of few tens of Myr only. To explain the observed features,
a more recent event is needed. 

The \ion{H}{i} distribution (Zhang et al.~\cite{zhang}) shows an unusual kink
in the eastern side close to the southern bar end. Heavily distorted radial
velocity and enhanced velocity dispersion is seen in the vicinity of this bar
end, also in other observations -- in \ion{H}{i}
(Zhang at al.~\cite{zhang}), H$\alpha$ (Chemin et al.~\cite{chemin}) and in CO
(e.g. Reuter et al.~\cite{reuter}; Dumke et al.~\cite{dumke}). In the last
case even a double line structure is clearly seen. 
The facts mentioned above fit the scenario of a recent collision with
a dwarf galaxy. The galaxy could enter the disk rouhgly along the eastern magnetic arm (Fig.~\ref{3627pi6}) 
and disturb the soutern end of the bar. This collision could induce locally enhanced star formation
and explain the existence of massive binary systems there, as well as the distortions of 
the southern bar end and the eastern spiral arm.
The admixture of colder dwarf matter can explain the lower thermal gas
temperature around the southern bar end seen in X-rays.

A recent collision also agrees with the unusual magnetic field configuration
in NGC\,3627 (Soida et al.~\cite{soida01}). Magnetohydrodynamics models of galactic magnetic
field evolution (e.g. Otmianowska \& Vollmer~\cite{otmian03}; Hanasz et al.~\cite{hanasz09}) show that a large-scale magnetic field needs
hundreds of Myr to develop and adopt to the large-scale gas flow. Because the
proposed collision took place a few tens of Myr ago only, the large-scale
magnetic field still ``remembers'' the pre-collision gas flow. 

In this scenario, a dwarf galaxy could enter the disk of NGC\,3627 from above and roughly along its eastern magnetic arm. We probably see the debris
of this dwarf (possibly its nucleus) as a distinct feature in the velocity
field seen in CO line observations (Reuter et al.~\cite{reuter}; Dumke et al.~\cite{dumke}).

\section{Summary and conclusions}
\label{cons}

We presented the analysis of the XMM-Newton X-ray data of NGC\,3627 and conclude the following.

\begin{itemize}
\item[-] In the northern bar end we found much hotter gas than in the southern one, which confirms the optically visible asymmetry, possibly caused by a tidal interaction
\item[-] An extended polarized radio ridge in the western arm of the galaxy is most likely caused by the ram-pressure effects within the Leo Triplet galaxy group, because we found 
	 signs for an increase in the temperature which, together with a gas density comparable to that of the disk, suggests compressions by the shock.
\item[-] A slightly asymmetric distribution of the hot gas in the eastern part of the central galactic disk follows the magnetic arm.
\item[-] To explain the observed features of NGC\,3627 we propose a rapid movement of the galaxy through the IGM and a 
	 collision with a companion dwarf galaxy entering the disk from the southeastern direction some tens of Myr ago. 
\end{itemize}

\begin{acknowledgements}
This work was supported by DLR Verbundforschung
"Extraterrestrische Physik" at the Ruhr-University Bochum through
grant 50 OR 0801. We thank Matthias Ehle, and the anonymous referee for valuable comments that helped to 
improve this paper.
\end{acknowledgements}


\begin{thebibliography}{}

\bibitem[1996]{xspec} Arnaud, K. A. 1996, Astronomical Data Analysis Software and Systems V, eds. Jacoby G. and Barnes J., p17, ASP Conf. Series volume 101

\bibitem[2011]{bogdan} Bogd\'an, \'A, \& Gilfanov, M. 2011, \mnras, 418, 1901

\bibitem[2011]{brightman11} Brightman, M., \& Nandra, K. 2011, \mnras, 413, 1206

\bibitem[2007]{carter} Carter, J. A., \& Read, A. M. 2007, \aap, 464, 1155

\bibitem[2003]{chemin} 	Chemin, L., Cayatte, V., Balkowski, C., et al. 2003, \aap, 405, 89

\bibitem[2011]{corral11} Corral, A., Della Ceca, R., Caccianiga, A., et al. 2011, \aap, 530, 42

\bibitem[2011]{dumke} Dumke, M., Krause, M., Beck, R., et al. 2011, ASPC, 446, 111 

\bibitem[2010]{evans10} Evans, I. N., Primini, F. A., Glotfelty, K. J., et al. 2010, \apjs, 189, 37

\bibitem[2004]{sas} Gabriel, C., Denby, M., Fyfe, D. J., et al. 2004, ASPC, 314, 759

\bibitem[2005]{galbiati05} Galbiati, E., Caccianiga, A., Maccacaro, T., et al. 2005, \aap, 430, 927

\bibitem[1972]{gunn} Gunn, J. E. \& Gott, J. R. 1972, \apj, 176, 1

\bibitem[2009]{hanasz09} Hanasz, M., Otmianowska-Mazur, K., Kowal, G., \& Lesch, H. 2004, \aap, 498, 335

\bibitem[1979]{haynes79} Haynes, M. P., Giovanelli, R., \& Roberts, M. S., 1979, \apj, 229, 83

\bibitem[2001]{jansen} Jansen, F., Lumb, D., Altieri, B., et al. 2001, \aap, 365, 1

\bibitem[1992]{kaastra} Kaastra, J. S. 1992, An X-Ray Spectral Code for Optically Thin Plasmas (Internal SRON-Leiden Report, updated version 2.0)

\bibitem[2005]{lab} Kalberla, P. M. W., Burton, W. B., Hartmann D., et al. 2005, \aap, 440, 775

\bibitem[2003]{kanbur03} Kanbur, S. M., Ngeow, C., Nikolaev, S., et al. 2003, \aap, 411, 361

\bibitem[2008]{sings} Kennicutt, R. C., Lee, J. C., Funes, J. G., et al. 2008, \apjs, 178, 247

\bibitem[1985]{mewe} Mewe, R., Gronenschild, E. H. B. M., \& van den Oord, G. H. J. 1985, \aaps, 62, 197

\bibitem[2004]{spcheck} Molendi, S., De Luca, A., \& Leccardi, A. 2004, \aap, 419, 837

\bibitem[2003]{otmian03} Otmianowska-Mazur, K., \& Vollmer, B., 2003, A\&A, 402, 879

\bibitem[2009]{owen09} Owen, R. A., \& Warwick, R. S. 2009, \mnras, 394, 1741

\bibitem[2008]{paladino} Paladino, R., Murgia, M., Tarchi, A., et al. 2008, \aap, 485, 679

\bibitem[1983]{leda} Paturel, G. 2003, \aap, 412, 45

\bibitem[1996]{reuter} Reuter, H.-P., Sievers, A.W, Pohl, M., et al. 1996, \aap, 306, 721

\bibitem[1978]{rots} Rots, A. H. 1978, \aj, 83, 3

\bibitem[2011]{rotsbuda} Rots, A. H., \& Budav\'ari, T. 2011, \apjs, 192, 8

\bibitem[2008]{saxton08} Saxton, R. D., Read, A. M., Esquej, P., et al. 2008, \aap, 480, 611

\bibitem[2001]{soida01} Soida, M., Urbanik, M., Beck, R., et al. 2001, \aap, 378, 40

\bibitem[2011]{swartz} Swartz, D. A., Soria, R., Tennant, A. F., \& Yukita, M. 2011, \apj, 741, 49

\bibitem[1998]{schlegel98} Schlegel, D. J., Finkbeiner, D. P., Davis, M. 1998, \apj, 500, 525

\bibitem[2001]{strueder} Str\"uder, L., Briel, U., Dennerl., K., et al. 2001, \aap, 365, 18

\bibitem[1972]{toomre} Toomre, A., \& Toomre, J. 1972, \apj, 178, 623

\bibitem[2001]{turner} Turner, M. J. L., Abbey, A., Arnaud, M., et al. 2001, \aap, 365, 27

\bibitem[2006]{tuellmann06} T\"ullmann, R., Pietch, W., Rossa, J., et al. 2006, \aap, 448, 43

\bibitem[2005]{vazquez05} V\'azquez, G. A. \& Leitherer, C. 2005, \apj, 621, 695

\bibitem[2011]{watanabe} Watanabe, Y., Sorai, K., Kuno, N., \& Habe, A. 2011, \mnras, 411, 1409

\bibitem[2010]{Zacharias2010} Zacharias, N., Finch, C, Girard, T, et al. 2010, \aj, 139, 2184

\bibitem[1993]{zhang} Zhang, X., Wright, M., \& Alexander, P. 1993, \apj, 418, 100

\end{thebibliography}
\end{document}